\documentclass[a4paper,twocolumn,showpacs,prl,amsmath,amssymb,superscriptaddress]{revtex4}
\pdfoutput=1
\usepackage{graphicx}% Include figure files
\usepackage{color}% Include colors
\usepackage{dcolumn}% Align table columns on decimal point
\usepackage{bm}% bold math

\begin{document}

\preprint{Ver. 0.5}

\title{Optimal generation of indistinguishable photons \\
from non-identical artificial molecules}

\author{Emiliano Cancellieri}
        \affiliation{NEST CNR-INFM and Scuola Normale Superiore, Pisa, Italy}
        \affiliation{S3 CNR-INFM, Modena, Italy}
\author{Filippo Troiani}
       \affiliation{S3 CNR-INFM, Modena, Italy}
       \email{filippo.troiani@unimore.it}
\author{Guido Goldoni}
       \affiliation{S3 CNR-INFM, Modena, Italy}
       \affiliation{Dipartimento di Fisica, Universit\`a di Modena e Reggio Emilia}

\date{\today}

\begin{abstract}
We  show theoretically that nearly indistinguishable photons can be generated with
non-identical semiconductor-based sources.
The use of virtual Raman transitions and the optimization of the external driving
fields increases the tolerance to spectral inhomogeneity to the meV energy range.
A trade-off emerges between photon indistinguishability and efficiency in the
photon-generation process.
Linear (quadratic) dependence of the coincidence probability within the
Hong-Ou-Mandel setup is found with respect to the dephasing (relaxation) rate
in the semiconductor sources.

\end{abstract}

\pacs{03.67.-a, 78.67.Hc, 42.50.-p}

\maketitle

Single-photon sources (SPSs) are fundamental devices in quantum communication
\cite{cirac97} and linear-optics quantum computation \cite{kok07}.
Essentially, a SPS consists of an atomic-like system that can be deterministically
excited and thus triggered to emit single-photon wavepackets into a preferential
mode.
In addition, all photons have to be emitted in the same quantum state, in order to
maximize the visibility of two-photon interference.

\begin{figure}
\includegraphics[width=\columnwidth]{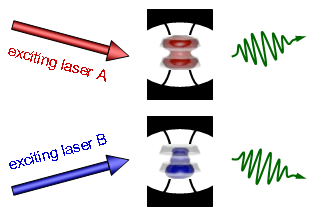}
\caption{\label{fig1} (Color online) Emission scheme of indistinguishable photons from non-identical
SPSs. Each source (A/B) consists of an AM (gray), doped with
an excess electron (red/blue), and coupled to an optical MC (black).
The effect on the photon wavepackets (green) of the differences between A and B,
in terms of energies and oscillator strengths of the optical transitions,
are compensated by properly tailoring the exciting laser pulses (straight arrows).}
\end{figure}

So far, single-photon generation has been demonstrated both with
atomic \cite{keller04,mckeever04,
darquie05,beugnon06}, molecular \cite{dmartini96, brunel99,lounis00},
and solid-state systems
\cite{kurtsiefer00,michler00,zwiller01,moreau01,santori01,santori}.
Amongst the latter, self-assembled semiconductor quantum dots (SAQDs) \cite{zrenner}
are particularly promising, for they
can be controllably coupled to optical microcavities \cite{badolato}, so as to
greatly enhance both the photon emission rate and the collection efficiency.
As a major downside, SAQDs come with a finite dispersion in terms of size,
shape, and composition. This typically gives rise to spectral inhomogeneities in
the meV range, {\it i.e.}, orders of magntitude larger than the homogeneous
linewidths (few $\mu $eVs) of the lowest exciton transitions at cryogenic temperatures.
Photons spontaneously emitted by two distinct SAQDs therefore tend to be completely
distinguishable: this could impede scalability and ultimately limit the potential
of semiconductor-based SPSs.

In spite of its relevance to any solid-state approach, the problem of generating
indistinguishable photons with non-identical emitters has still received limited
attention.
In this Letter, we assess the possibility of compensating the effects of such
differences between two SPSs by suitably tuning the exciting laser pulses.
To this aim, we combine a density-matrix approach  \cite{scully} -- to simulate the system
dynamics and compute the coherence functions of the emitted radiation --
with a genetic algorithm \cite{michalewicz} -- to optimize the external
driving fields.
As a crucial point, the photon generation results from a (virtual) Raman transition.
Raman transitions have already been proposed as means to avoid the classical
uncertainty on the initial time of the emission process (the so-called {\it time-jitter})
 \cite{kiraz04,troiani06a} and to tune both the temporal profile
and the central frequency of the emitted wavepacket \cite{mckeever04}.
Here we show that such flexibility can be exploited to generate two nearly
indistinguishable photons, in spite of the spectral differences between their
respective semiconductor sources.

\begin{figure}
\includegraphics[width=\columnwidth]{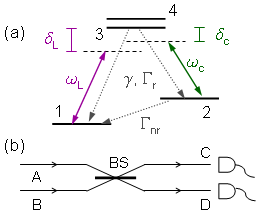}
\caption{\label{fig2}
(Color online) (a) Relevant level scheme for each negatively-charged AM.
The levels 1 and 2 correspond to the bonding and antibonding states of the
excess electron; these are optically coupled to the two lowest charged
exciton states 3 and 4.
The off-resonant laser pulse ($\omega_L$) induces a virtual Raman transition
from 1 to 2; this process results in the creation and emission of a photon
from the fundamental mode of the MC ($\omega_c$).
(b) Schematics of the Hong-Ou-Mandel setup.
The photons are emitted by the two sources A and B in the input
modes of a balanced beam-splitter (BS), whose output modes are
coupled to the photodetectors C and D.}
\end{figure}
%
% {\it Method ---}
We specifically consider the case where each source is represented by two
coherently coupled SAQDs (often referred to as {\it artificial molecule}, AM),
embedded in
an optical microcavity (MC) (Fig. \ref{fig1}).
The AM is doped with an excess electron, whose levels 1 and 2
[Fig. \ref{fig2}(a)] define a {\it pseudo-spin},
corresponding to the hybridized -- bonding and antibonding --
states of the two dots \cite{sheng02,bester04};
together with the lowest charged exciton states (3 and 4), these form a double
$\Lambda -$scheme \cite{nota_a2}.
There, a Raman transition can be induced by an off-resonant laser pulse
($\omega_L$), that drives the AM from 1 to 2, while generating a cavity photon
($\omega_c$).

We simulate the dynamics of the two sources ($A$ and $B$) within a
density-matrix approach \cite{scully}.
The state of each AM-MC system is denoted by $ | j , n \rangle $, where
$ j = 1, \dots , 4 $ specifies the AM eigenstate, whereas $ n $ represents
the occupation number of the cavity mode.
The time evolution of the density matrix within such Hilbert space is given
by
$ \dot{\rho} = i [\rho , H_{MC} + H_{L}] + {\mathcal L} \rho $
(rotating-wave approximation, $\hbar = 1$).
The coupling of the AM with the MC is accounted by
\begin{equation}
H_{MC} = \sum_{j=1,2} \sum_{k=3,4} g_{jk} (\sigma_{jk} a^{\dagger} +
a \sigma_{jk}^\dagger) ,
\end{equation}
where $a$ is the annihilation operator of the cavity photon and
$ \sigma_{jk} \equiv | j \rangle\langle k | $ the ladder operator acting
on the AM state.
The coupling of the AM with the exciting laser is given by
\begin{equation}\label{eq2}
H_{L} = \frac{1}{2} \sum_{p=1}^{N_L} \sum_{j=1,2} \sum_{k=3,4} \Omega^{p}_{jk} (t)
({\rm e}^{-i\delta_{jk}^p t}\sigma_{jk} + {\rm H. c.} ) ,
\end{equation}
where
$ \Omega^{p}_{jk} (t) / g_{jk} = \Omega_{0}^p \exp \{(t-t_{0}^p)^2 /
2 \sigma_p^2\}$
gives the time envelope of the $p-$th laser pulse,
$ \delta_{jk}^p \equiv \epsilon_k - \epsilon_j - \omega_L^p $,
and $\epsilon_j$ the AM levels.
The interaction of the AM-MC system with the environment is given by the
superoperator \begin{eqnarray}
\label{eq1}
{\mathcal L}
=
\sum_{j} L (\sqrt{\gamma_j}P_j)
+ \sum_{j,k} L (\sqrt{\Gamma_{jk}}\sigma_{jk})
+ L (\sqrt{\kappa} a)  ,
\end{eqnarray}
where
$ L(A)\rho \equiv (2A \rho A^\dagger - A^\dagger A \rho - \rho A^\dagger A)/2$
and
$ P_j \equiv | j \rangle\langle j | $.
The three terms in Eq. \ref{eq1} account for pure dephasing
(with rate $\gamma$),
radiative ($ jk = 13,14,23,24 $) and non-radiative
($jk=12, 34$)
relaxation of the AM, and cavity-photon emission, respectively.
In the following, we assume for simplicity that $\Gamma_{jk}\equiv\Gamma_r$
for all the radiative relaxation processes of the AM, and
$\Gamma_{jk} \equiv \Gamma_{nr}$ for the non-radiative ones.

The degree of indistinguishability between the photons generated by
the two sources can be measured within the Hong-Ou-Mandel setup \cite{hong}
 [Fig. \ref{fig2}(b)].
There, if two indistinguishable photons enter the input ports (A and B) of
a balanced beam splitter, two-photon interference results in a vanishing
probability of a coincidence event ($P_{CD}=0$) in the two photodetectors at
the output modes (C and D). For distinguishable photons,
instead, $P_{CD}=1/2$.
The coincidence probability can thus be regarded as a measure of the
photon indistinguishability.
Formally, $P_{CD}$ can be connected to the dynamics of the $A$ and $B$
sources through \cite{kiraz04}
$
P_{CD} = F_1 - F_2 ,
$
where
$ F_{p=1,2} = \kappa^2 \int dt\int d\tau\, f_p (t,\tau) $
and
\begin{eqnarray*}
f_1 (t,\tau) &=& [ n_A (t+\tau) n_B(t) + n_B(t) n_A (t+\tau)] / 4 ,
\\
f_2 (t,\tau) &=&
{\rm Re}
\left\{ [ G^{(1)}_A(t,t+\tau) ]^* \, G^{(1)}_B(t,t+\tau) \right\} / 2.
\end{eqnarray*}
Here,
$ G_\chi^{(1)} (t,t+\tau) = \langle a^{\dagger}_\chi (t)a_\chi (t+\tau)\rangle $
and
$ n_\chi (t) = \langle a_\chi^\dagger (t) a_\chi (t) \rangle $
are, respectively, the first-order coherence function and cavity-mode
occupation corresponding to the sources $\chi = A,B$.

In order to maximize the overlap between the wavepackets of the photons
emitted by A and B, we optimize the driving laser pulses
and the frequency of the cavity mode, {\it i.e.}, the vector
$ {\bf X} = ( \Omega_0^n , t_0^n , \sigma_n , \omega_c ) $.
For two given sources [each characterized by the vector
$ {\bf Y} = ( \epsilon_j , g_{jk} , \gamma_j , \Gamma_{jk} ) $,
with $ {\bf Y}_A \neq {\bf Y}_B $],
the suitability of each set of laser pulses for the generation of two
indistinguishable photons is quantified by a fitness function
$ \mathcal{F} ( {\bf X}_A , {\bf X}_B | {\bf Y}_A , {\bf Y}_B ) \ge 0 $,
defined as
\begin{equation}
\mathcal{F} = ( F_2 / F_1 ) \, g ( P_e^A , P_e^B ) .
\end{equation}
Here, $ 0 \le F_2/F_1 \le 1 $ measures the degree of indistinguishability
between the two photons, whereas $g$ accounts for the statistics of the two
SPSs; more specifically, it imposes a penalty on the vectors $ {\bf X}_\chi $
corresponding to a photon-emission probability
$ P_e^\chi = \int n_\chi (t) dt $ far from 1.
% Essentially, the genetic algorithm proceeds as follows:
% initially, a pool of $N_P$ individuals is randomly generated, each individual
% being identified by the vector ${\bf X}_{AB} \equiv ({\bf X}_A,{\bf X}_B)$.
% The population is evolved by first generating new individuals, resulting from
% the application to $ {\bf X}_{AB} $ of unary and binary operators, and
% subsequently selecting the fittest $N_P$ individuals according to the values
% of $\mathcal{F}$.
% After iterating such process a finite number $N_G$ of times,
We identify the best solution ($ {\bf X}_\chi^0 $) with the vectors
that correspond to the maximum value of the fitness function, for the given
values of the physical parameters that characterize the SPSs ($ {\bf Y}_\chi ) $:
$ \mathcal{F}_M = \mathcal{F} ( {\bf X}_A^0 , {\bf X}_B^0 | {\bf Y}_A ,
{\bf Y}_B ) $.
Hereafter, we take:
$ g = \theta (  P_e^A - P_e^t ) \cdot \theta (  P_e^B - P_e^t ) $,
with $ \theta $ the Heaviside function.
Therefore, provided that both $P_e^A$ and $P_e^B$ are larger than the
threshold $P_e^t$, $1-\mathcal{F}$ coincides with the coincidence probability
$P_{CD}$, normalized to the joint emission probability of the two sources
$P_{AB}=P_e^A \cdot P_e^B=F_1$.
To this aim, we combine the density matrix approach with a genetic algorithm
\cite{michalewicz}; this allows to efficiently explore, within a large
parameter space, the particularly complex landscape induced by the penalty function $g$.

\begin{figure}
\includegraphics[width=0.8\columnwidth]{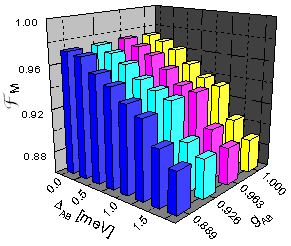}
\caption{\label{fig3}
(Color online) Maximized fitness function $ \mathcal{F}_M $ as a function of the differences
between the A and B sources in terms of frequencies ($ \Delta_{AB} = \omega^A_{32}
- \omega^B_{32} $) and oscillator strengths ($ g_{AB} \equiv g^B_{jk} / g^A_{jk} $)
of the optical transitions.
The remaining parameters are:
$ \Gamma_{nr} = \Gamma_r^{A/B} = 10^{-3} \,$ps$^{-1}$, $\gamma^{A/B}_j=10^{-2}\,$ps$^{-1}$,
$ \kappa^{A/B} = 10^{-1}\, $ps$^{-1}$, $ g^A_{jk} = 0.120\,$meV.
Having found no evidence of an increased indistinguishability arising from
non-gaussian laser-pulse profiles, we set here $ N_L = 1$.}
\end{figure}

% {\it Results ---}
The distinguishability between the photons emitted by the two sources
mainly arises from the differences between their transition energies
($\omega_{jk}^A\neq\omega_{jk}^B$, being $\omega_{jk}\equiv\epsilon_j-\epsilon_k$)
and oscillator strengths ($ g_{jk}^A \neq g_{jk}^B$).
In Fig. \ref{fig3}
we show the dependence of the maximum fitness for optimized laser pulses,
$ \mathcal{F}_M $,
on
$ g_{AB} \equiv g^A_{jk} / g^B_{jk} $
and
$ \Delta_{AB} = \omega_{32}^A - \omega_{32}^B $ \cite{nota_a1},
for a threshold emission probability $P_e^t = 0.9$.
% being
% $ \delta_c^\chi = \omega^\chi_{32} - \omega_c $
% the detuning between the energy of the $ |3\rangle \rightarrow |2\rangle $
% transition and that of the cavity photon [Fig. \ref{fig2}(a)].
% The optimized value of $\omega_c$ typically falls in between
% $\omega_{32}^A$ and $\omega_{32}^B$, such that $\delta_c^B\,\delta_c^B < 0 $;
% $|\Delta_{AB}|$ thus quantifies the overall detuning of the emitters from the
% MC mode.
% As result of an optimization process, the obtained values for $P^0_{CD}$,
% must be considered as a higher limit for the theoretical minimal value for
% $P^0_{CD}$. Owing to the discrete nature of our numerical simulation we estimate
% an error of about $0.005$ on the best obtained value for $P^0_{CD}$.
The fact that $ \mathcal{F}_M < 1 $ even for the case of identical
sources ($g_{AB}=1$ and $\Delta_{AB}=0$) is due to the presence of dephasing
and non-radiative relaxation (see below).
By adjusting the laser pulses, moderate differences between the oscillator 
strengths of the two sources ($g_{AB} \gtrsim 0.9 $) can be efficiently 
compensated, whereas decreases of $\mathcal{F}_M$ arising from 
mismatches between the optical-transition energies of A and B in the meV
range can be limited to a few percent.

A couple of general comments are in order.
On the one hand, stimulated Raman processes do increase the tolerance to
inhomogeneities between the SPSs from the $\mu$eV -- as is the case with spontaneous
emission -- to the meV range.
On the other hand, such inhomogeneities cannot be completely compensated by
properly tuning the frequencies of the driving laser pulses, as could be naively
expected on the basis of a simple energy-conservation relation.
These limitations are partially due to the effects of decoherence; besides, the
requirement that the two sources emit photons with a high probability seems to
conflict with that of maximizing their indistinguishability.
In the following we investigate separately these two aspects.

\begin{figure}
\includegraphics[width=\columnwidth]{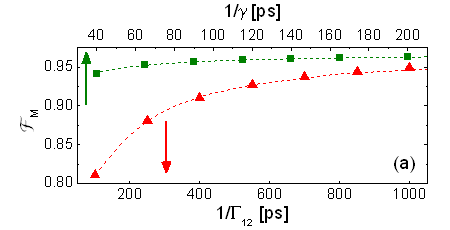}
\includegraphics[width=\columnwidth]{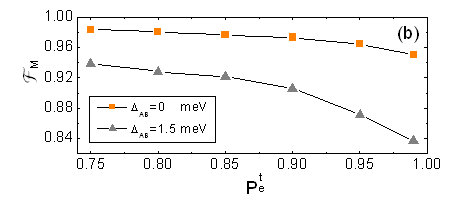}
\caption{\label{fig4} (Color online)
(a) Dependence of the maximum fitness function $\mathcal{F}_M$ on the
dephasing rate $\gamma$ (upper axis, green squares) and on the non-radiative
relaxation rate $\Gamma_{21}$ (lower axis, red triangles),
for $g^{AB}=1$ and $\Delta_{AB}=1.25$ meV.
The green (red) dotted line represents the best fit of $\mathcal{F}_M$
linear (quadratic) in $\gamma$ ($\Gamma_{21}$).
(b) $\mathcal{F}_M$ as a function of the threshold $P_e^t$ in the
photon-emission probability for $g_{AB} = 1$, with $ \Delta_{AB}=0 $
and $1.5$ meV. }
\end{figure}

The coupling between the carriers confined in each AM and the phonons gives
rise to energy relaxation and dephasing.
These two contributions are considered separately in Fig. \ref{fig4}(a), where
we report $\mathcal{F}_M$ as a function of $ \gamma $ (with $ \Gamma_{21} = 0$,
squares), and of $ \Gamma_{21} $ (with $ \gamma = 0 $, triangles),
within realistic ranges of values for these rates.
As in the case of two photons sequentially emitted by the same source
\cite{kiraz04},
the coincidence probability increases linearly with $ \gamma $
(dotted green curve).
A stronger dependence is found with respect to the non-radiative relaxation
between the two lowest states in the $\Lambda -$scheme (a fit by a quadratic
function of $\Gamma_{21}$ is given by the dotted red line).
We note, however, that the value of $\Gamma_{21}$ can be strongly reduced by
suitably engineering the AM geometry and by the application of external fields
\cite{zanardi,bertoni}.

In order to investigate the relation between the emission probability of
the A and B photons and their indistinguishability, we have computed
$\mathcal{F}_M$ as a function of $P_e^t$.
In Fig. \ref{fig4}(b) we report two representative cases: $\Delta_{AB} =
0$ (orange squares) and $\Delta_{AB}=1.5\,$meV (gray triangles), with $g_{AB}=1$.
The photon emission probability and the degree of indistinguishability are
clearly anti-correlated, thus demonstrating the existence of a trade-off between
the two requirements.
The dependence of $\mathcal{F}_M$ on $P_e^t$, though evident even for identical
sources ($\Delta_{AB}=0$), is more pronounced in the presence of a spectral
mismatch ($\Delta_{AB}=1.5$).
The above curves are relatively insensitive to differences in the oscillator
strengths ($g_{AB} \gtrsim 0.9 $, not shown here).

\begin{figure}
\includegraphics[width=\columnwidth]{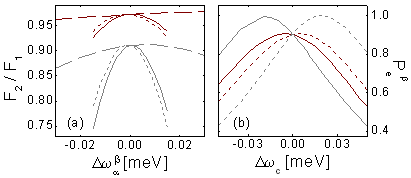}
\caption{\label{fig5} (Color online) (a) Dependence of
$F_2/F_1 = 1 - P_{CD}$ on the shift of the cavity and laser frequencies from their
optimized values,
$\Delta\omega_\alpha^{A/B} \equiv \omega_\alpha^{A/B} - \omega_\alpha^{A/B,0}$,
for $\Delta_{AB}=0$ (red) and $\Delta_{AB}=1.5\,$meV (gray), with $g_{AB}=0.88$.
Solid, dashed and dotted lines correspond respectively to:
$\alpha = c$; $\alpha = L$, $\beta = A$; $\alpha = L$, $\beta = B$.
(b) Photon-emission probabilities, $P_e^{A}$ (solid lines) and $P_e^{B}$ (dotted),
as a function of $\Delta\omega_c$, for $\Delta_{AB}=0$ (red) and $\Delta_{AB}=1.5\,$meV
(gray), with $g_{AB}=0.88$.}
\end{figure}

We finally comment on the robustness of the solution with respect to small
departures of the control parameters from their optimal values.
As a representative example, we report the dependence of $ F_2 / F_1 = 1 - P_{CD} $
on the cavity frequency $ \omega_c \equiv \omega_c^A = \omega_c^B $, referred to its optimized
value $\omega_c^0$ [Fig. \ref{fig5} (a)].
The robustness of the photon indistinguishability with respect to non-optimal
parameters decreases for incresing spectral mismatches between the two sources.
Besides, our simulations show a stronger dependence of $ P_{CD} $ on the laser
(solid and dotted lines) than on the cavity frequencies (dashed).
This feature can be traced to the fact that the uncertainty on the cavity
frequency ($\sim 1/\kappa = 10\,$ps) is larger than that on the laser frequency
(typical duration of the optimized laser pulses are few tens of ps).
As to $P_e^{A/B}$ [Fig. \ref{fig5} (b)], we note that the value corresponding
to the optimized parameter ($\Delta\omega_c =0$) is close to the mimimum allowed
value ($P_e^t$): this provides a further indication on the existence of a
trade-off between photon indistinguishability and emission probability.
Besides, the existence of a spectral mismatch (gray lines) makes also the
fulfilment of the requirement $ P_e^A, P_e^B > P_e^t $ more sensitive to the
tuning of the physical parameters. The required precision is of the order of
a few $\mu$eV for $\Delta_{AB}=1.5\,$meV.
%

% {\it Conclusions ---}
In conclusion, we have investigated the degree of indistinguishability
between single photons emitted by non-identical artificial molecules.
We found that the use of virtual Raman transitions, combined with the
optimal design of the driving laser pulses, increases the tolerance with
respect to the spectral inhomogeneities between the sources up to the
meV energy range. Defined power-law dependences of the coincidence
probability on the dephasing and (non-radiative) relaxation rates are
identified. Besides, unlike the case of identical sources, a trade-off
emerges between the requirements of maximizing the emission efficiency
and the photon indistinguishability.
It is finally worth mentioning that the use of a pseudo-spin within the
$\Lambda$-level scheme offers promising possibilities \cite{troiani07}.
These include the use of quantum-confined Stark effect \cite{krenner06}
for triggering the photon emission \cite{fernee07}, and the
entangling of carrier spins localized in remote (and thus non-identical)
semiconductor systems \cite{flindt}. 
The latter goal requires the combination of a
spin-photon entangling \cite{cortez} process with two-photon
interference within the Hong-Ou-Mandel setup.

We acknowledge financial support from the Italian MIUR under FIRB Contract
No. RBIN01EY74.


\begin{thebibliography}{31}
\expandafter\ifx\csname natexlab\endcsname\relax\def\natexlab#1{#1}\fi
\expandafter\ifx\csname bibnamefont\endcsname\relax
  \def\bibnamefont#1{#1}\fi
\expandafter\ifx\csname bibfnamefont\endcsname\relax
  \def\bibfnamefont#1{#1}\fi
\expandafter\ifx\csname citenamefont\endcsname\relax
  \def\citenamefont#1{#1}\fi
\expandafter\ifx\csname url\endcsname\relax
  \def\url#1{\texttt{#1}}\fi
\expandafter\ifx\csname urlprefix\endcsname\relax\def\urlprefix{URL }\fi
\providecommand{\bibinfo}[2]{#2}
\providecommand{\eprint}[2][]{\url{#2}}

\bibitem[{\citenamefont{Cirac et~al.}(1997)\citenamefont{Cirac, Zoller, Kimble,
  and Mabuchi}}]{cirac97}
\bibinfo{author}{\bibfnamefont{J.~I.} \bibnamefont{Cirac}},
  \bibinfo{author}{\bibfnamefont{P.}~\bibnamefont{Zoller}},
  \bibinfo{author}{\bibfnamefont{H.~J.} \bibnamefont{Kimble}},
  \bibnamefont{and} \bibinfo{author}{\bibfnamefont{H.}~\bibnamefont{Mabuchi}},
  \bibinfo{journal}{Phys. Rev. Lett.} \textbf{\bibinfo{volume}{78}},
  \bibinfo{pages}{3221} (\bibinfo{year}{1997}).

\bibitem[{\citenamefont{Kok et~al.}(2007)\citenamefont{Kok, Munro, Nemoto,
  Ralph, Dowling, and Milburn}}]{kok07}
\bibinfo{author}{\bibfnamefont{P.}~\bibnamefont{Kok}},
  \bibinfo{author}{\bibfnamefont{W.~J.} \bibnamefont{Munro}},
  \bibinfo{author}{\bibfnamefont{K.}~\bibnamefont{Nemoto}},
  \bibinfo{author}{\bibfnamefont{T.~C.} \bibnamefont{Ralph}},
  \bibinfo{author}{\bibfnamefont{J.~P.} \bibnamefont{Dowling}},
  \bibnamefont{and} \bibinfo{author}{\bibfnamefont{G.~J.}
  \bibnamefont{Milburn}}, \bibinfo{journal}{Rev. Mod. Phys.}
  \textbf{\bibinfo{volume}{79}}, \bibinfo{pages}{135} (\bibinfo{year}{2007}).

\bibitem[{\citenamefont{Keller et~al.}(2004)\citenamefont{Keller, Lange,
  Hayasaka, Lange, and Walther}}]{keller04}
\bibinfo{author}{\bibfnamefont{M.}~\bibnamefont{Keller}},
  \bibinfo{author}{\bibfnamefont{B.}~\bibnamefont{Lange}},
  \bibinfo{author}{\bibfnamefont{K.}~\bibnamefont{Hayasaka}},
  \bibinfo{author}{\bibfnamefont{W.}~\bibnamefont{Lange}}, \bibnamefont{and}
  \bibinfo{author}{\bibfnamefont{H.}~\bibnamefont{Walther}},
  \bibinfo{journal}{Nature} \textbf{\bibinfo{volume}{431}},
  \bibinfo{pages}{1075} (\bibinfo{year}{2004}).

\bibitem[{\citenamefont{McKeever et~al.}(2004)\citenamefont{McKeever, Boca,
  Boozer, Miller, Miller, Buck, Kuzmich, and Kimble}}]{mckeever04}
\bibinfo{author}{\bibfnamefont{J.}~\bibnamefont{McKeever}},
  \bibinfo{author}{\bibfnamefont{A.}~\bibnamefont{Boca}},
  \bibinfo{author}{\bibfnamefont{A.~D.} \bibnamefont{Boozer}},
  \bibinfo{author}{\bibfnamefont{R.}~\bibnamefont{Miller}},
  \bibinfo{author}{\bibfnamefont{J.~R.} \bibnamefont{Miller}},
  \bibinfo{author}{\bibfnamefont{J.~R.} \bibnamefont{Buck}},
  \bibinfo{author}{\bibfnamefont{A.}~\bibnamefont{Kuzmich}}, \bibnamefont{and}
  \bibinfo{author}{\bibfnamefont{H.~J.} \bibnamefont{Kimble}},
  \bibinfo{journal}{Science} \textbf{\bibinfo{volume}{303}},
  \bibinfo{pages}{1992} (\bibinfo{year}{2004}).

\bibitem[{\citenamefont{Darqui\'e et~al.}(2005)\citenamefont{Darqui\'e, Jones,
  Dingjan, Beugnon, Bergamini, Sortais, Messin, Browaeys, and
  Grangier}}]{darquie05}
\bibinfo{author}{\bibfnamefont{B.}~\bibnamefont{Darqui\'e}},
  \bibinfo{author}{\bibfnamefont{M.~P.~A.} \bibnamefont{Jones}},
  \bibinfo{author}{\bibfnamefont{J.}~\bibnamefont{Dingjan}},
  \bibinfo{author}{\bibfnamefont{J.}~\bibnamefont{Beugnon}},
  \bibinfo{author}{\bibfnamefont{S.}~\bibnamefont{Bergamini}},
  \bibinfo{author}{\bibfnamefont{Y.}~\bibnamefont{Sortais}},
  \bibinfo{author}{\bibfnamefont{G.}~\bibnamefont{Messin}},
  \bibinfo{author}{\bibfnamefont{A.}~\bibnamefont{Browaeys}}, \bibnamefont{and}
  \bibinfo{author}{\bibfnamefont{P.}~\bibnamefont{Grangier}},
  \bibinfo{journal}{Science} \textbf{\bibinfo{volume}{309}},
  \bibinfo{pages}{454} (\bibinfo{year}{2005}).

\bibitem[{\citenamefont{Beugnon et~al.}(2006)\citenamefont{Beugnon, Jones,
  Dingjan, Darqui\'e, Messin, Browaeys, and Grangier}}]{beugnon06}
\bibinfo{author}{\bibfnamefont{J.}~\bibnamefont{Beugnon}},
  \bibinfo{author}{\bibfnamefont{M.~P.~A.} \bibnamefont{Jones}},
  \bibinfo{author}{\bibfnamefont{J.}~\bibnamefont{Dingjan}},
  \bibinfo{author}{\bibfnamefont{B.}~\bibnamefont{Darqui\'e}},
  \bibinfo{author}{\bibfnamefont{G.}~\bibnamefont{Messin}},
  \bibinfo{author}{\bibfnamefont{A.}~\bibnamefont{Browaeys}}, \bibnamefont{and}
  \bibinfo{author}{\bibfnamefont{P.}~\bibnamefont{Grangier}},
  \bibinfo{journal}{Nature} \textbf{\bibinfo{volume}{440}},
  \bibinfo{pages}{779} (\bibinfo{year}{2006}).

\bibitem[{\citenamefont{Martini et~al.}(1996)\citenamefont{Martini, Giuseppe,
  and Marrocco}}]{dmartini96}
\bibinfo{author}{\bibfnamefont{F.~D.} \bibnamefont{Martini}},
  \bibinfo{author}{\bibfnamefont{G.~D.} \bibnamefont{Giuseppe}},
  \bibnamefont{and} \bibinfo{author}{\bibfnamefont{M.}~\bibnamefont{Marrocco}},
  \bibinfo{journal}{Phys. Rev. Lett.} \textbf{\bibinfo{volume}{76}},
  \bibinfo{pages}{900} (\bibinfo{year}{1996}).

\bibitem[{\citenamefont{Brunel et~al.}(1999)\citenamefont{Brunel, Lounis,
  Tamarat, and Orrit}}]{brunel99}
\bibinfo{author}{\bibfnamefont{C.}~\bibnamefont{Brunel}},
  \bibinfo{author}{\bibfnamefont{B.}~\bibnamefont{Lounis}},
  \bibinfo{author}{\bibfnamefont{P.}~\bibnamefont{Tamarat}}, \bibnamefont{and}
  \bibinfo{author}{\bibfnamefont{M.}~\bibnamefont{Orrit}},
  \bibinfo{journal}{Phys. Rev. Lett.} \textbf{\bibinfo{volume}{83}},
  \bibinfo{pages}{2722} (\bibinfo{year}{1999}).

\bibitem[{\citenamefont{Lounis and Moerner}(2000)}]{lounis00}
\bibinfo{author}{\bibfnamefont{B.}~\bibnamefont{Lounis}} \bibnamefont{and}
  \bibinfo{author}{\bibfnamefont{W.~E.} \bibnamefont{Moerner}},
  \bibinfo{journal}{Nature} \textbf{\bibinfo{volume}{407}},
  \bibinfo{pages}{491} (\bibinfo{year}{2000}).

\bibitem[{\citenamefont{Kurtsiefer et~al.}(2000)\citenamefont{Kurtsiefer,
  Mayer, Zarda, and Weinfurter}}]{kurtsiefer00}
\bibinfo{author}{\bibfnamefont{C.}~\bibnamefont{Kurtsiefer}},
  \bibinfo{author}{\bibfnamefont{S.}~\bibnamefont{Mayer}},
  \bibinfo{author}{\bibfnamefont{P.}~\bibnamefont{Zarda}}, \bibnamefont{and}
  \bibinfo{author}{\bibfnamefont{H.}~\bibnamefont{Weinfurter}},
  \bibinfo{journal}{Phys. Rev. Lett.} \textbf{\bibinfo{volume}{85}},
  \bibinfo{pages}{290} (\bibinfo{year}{2000}).

\bibitem[{\citenamefont{Michler et~al.}(2000)\citenamefont{Michler, Imamoglu,
  Mason, Carson, Strouse, and Buratto}}]{michler00}
\bibinfo{author}{\bibfnamefont{P.}~\bibnamefont{Michler}},
  \bibinfo{author}{\bibfnamefont{A.}~\bibnamefont{Imamoglu}},
  \bibinfo{author}{\bibfnamefont{M.~D.} \bibnamefont{Mason}},
  \bibinfo{author}{\bibfnamefont{P.~J.} \bibnamefont{Carson}},
  \bibinfo{author}{\bibfnamefont{G.}~\bibnamefont{Strouse}}, \bibnamefont{and}
  \bibinfo{author}{\bibfnamefont{S.}~\bibnamefont{Buratto}},
  \bibinfo{journal}{Nature} \textbf{\bibinfo{volume}{406}},
  \bibinfo{pages}{968} (\bibinfo{year}{2000}).

\bibitem[{\citenamefont{Zwiller et~al.}(2001)\citenamefont{Zwiller, Blom,
  Jonsson, Panev, Jeppesen, Tsegaye, Goobar, Pistol, Samuelson, and
  Bjork}}]{zwiller01}
\bibinfo{author}{\bibfnamefont{V.}~\bibnamefont{Zwiller}},
  \bibinfo{author}{\bibfnamefont{H.}~\bibnamefont{Blom}},
  \bibinfo{author}{\bibfnamefont{P.}~\bibnamefont{Jonsson}},
  \bibinfo{author}{\bibfnamefont{N.}~\bibnamefont{Panev}},
  \bibinfo{author}{\bibfnamefont{S.}~\bibnamefont{Jeppesen}},
  \bibinfo{author}{\bibfnamefont{T.}~\bibnamefont{Tsegaye}},
  \bibinfo{author}{\bibfnamefont{E.}~\bibnamefont{Goobar}},
  \bibinfo{author}{\bibfnamefont{M.-E.} \bibnamefont{Pistol}},
  \bibinfo{author}{\bibfnamefont{L.}~\bibnamefont{Samuelson}},
  \bibnamefont{and} \bibinfo{author}{\bibfnamefont{G.}~\bibnamefont{Bjork}},
  \bibinfo{journal}{Appl. Phys. Lett.} \textbf{\bibinfo{volume}{78}},
  \bibinfo{pages}{2476} (\bibinfo{year}{2001}).

\bibitem[{\citenamefont{Moreau et~al.}(2001)\citenamefont{Moreau, Robert,
  Manin, Thierry-Mieg, Gerard, and Abram}}]{moreau01}
\bibinfo{author}{\bibfnamefont{E.}~\bibnamefont{Moreau}},
  \bibinfo{author}{\bibfnamefont{I.}~\bibnamefont{Robert}},
  \bibinfo{author}{\bibfnamefont{L.}~\bibnamefont{Manin}},
  \bibinfo{author}{\bibfnamefont{V.}~\bibnamefont{Thierry-Mieg}},
  \bibinfo{author}{\bibfnamefont{J.~M.} \bibnamefont{Gerard}},
  \bibnamefont{and} \bibinfo{author}{\bibfnamefont{I.}~\bibnamefont{Abram}},
  \bibinfo{journal}{Phys. Rev. Lett.} \textbf{\bibinfo{volume}{87}},
  \bibinfo{pages}{183601} (\bibinfo{year}{2001}).

\bibitem[{\citenamefont{Santori et~al.}(2001)\citenamefont{Santori, Pelton,
  Solomon, Dale, and Yamamoto}}]{santori01}
\bibinfo{author}{\bibfnamefont{C.}~\bibnamefont{Santori}},
  \bibinfo{author}{\bibfnamefont{M.}~\bibnamefont{Pelton}},
  \bibinfo{author}{\bibfnamefont{G.}~\bibnamefont{Solomon}},
  \bibinfo{author}{\bibfnamefont{Y.}~\bibnamefont{Dale}}, \bibnamefont{and}
  \bibinfo{author}{\bibfnamefont{Y.}~\bibnamefont{Yamamoto}},
  \bibinfo{journal}{Phys. Rev. Lett.} \textbf{\bibinfo{volume}{86}},
  \bibinfo{pages}{1502} (\bibinfo{year}{2001}).

\bibitem[{\citenamefont{Santori et~al.}(2002)\citenamefont{Santori, Fattal,
  Vuckovic, Solomon, and Yamamoto}}]{santori}
\bibinfo{author}{\bibfnamefont{C.}~\bibnamefont{Santori}},
  \bibinfo{author}{\bibfnamefont{D.}~\bibnamefont{Fattal}},
  \bibinfo{author}{\bibfnamefont{J.}~\bibnamefont{Vuckovic}},
  \bibinfo{author}{\bibfnamefont{G.}~\bibnamefont{Solomon}}, \bibnamefont{and}
  \bibinfo{author}{\bibfnamefont{Y.}~\bibnamefont{Yamamoto}},
  \bibinfo{journal}{Nature} \textbf{\bibinfo{volume}{419}},
  \bibinfo{pages}{594} (\bibinfo{year}{2002}).

\bibitem[{\citenamefont{Zrenner}(2000)}]{zrenner}
\bibinfo{author}{\bibfnamefont{A.}~\bibnamefont{Zrenner}}, \bibinfo{journal}{J.
  Chem. Phys.} \textbf{\bibinfo{volume}{112}}, \bibinfo{pages}{7790}
  (\bibinfo{year}{2000}).

\bibitem[{\citenamefont{Badolato et~al.}(2005)\citenamefont{Badolato, Hennessy,
  Atature, Dreyser, Hu, Petroff, and Imamoglu}}]{badolato}
\bibinfo{author}{\bibfnamefont{A.}~\bibnamefont{Badolato}},
  \bibinfo{author}{\bibfnamefont{K.}~\bibnamefont{Hennessy}},
  \bibinfo{author}{\bibfnamefont{M.}~\bibnamefont{Atature}},
  \bibinfo{author}{\bibfnamefont{J.}~\bibnamefont{Dreyser}},
  \bibinfo{author}{\bibfnamefont{E.}~\bibnamefont{Hu}},
  \bibinfo{author}{\bibfnamefont{P.~M.} \bibnamefont{Petroff}},
  \bibnamefont{and} \bibinfo{author}{\bibfnamefont{A.}~\bibnamefont{Imamoglu}},
  \bibinfo{journal}{Science} \textbf{\bibinfo{volume}{308}},
  \bibinfo{pages}{1158} (\bibinfo{year}{2005}).

\bibitem[{\citenamefont{Scully and Zubairy}(1997)}]{scully}
\bibinfo{author}{\bibfnamefont{M.~O.} \bibnamefont{Scully}} \bibnamefont{and}
  \bibinfo{author}{\bibfnamefont{M.}~\bibnamefont{Zubairy}}, in
  \emph{\bibinfo{booktitle}{Quantum optics}} (\bibinfo{publisher}{Cambridge
  University Press, Cambridge}, \bibinfo{year}{1997}).

\bibitem[{\citenamefont{Michalewicz}(1992)}]{michalewicz}
\bibinfo{author}{\bibfnamefont{Z.}~\bibnamefont{Michalewicz}}, in
  \emph{\bibinfo{booktitle}{Genetic Algorithms + Data Structures = Evolutionary
  Programming}} (\bibinfo{publisher}{Springer-Verlag, Berlin},
  \bibinfo{year}{1992}).

\bibitem[{\citenamefont{Kiraz et~al.}(2004)\citenamefont{Kiraz, Atature, and
  Imamoglu}}]{kiraz04}
\bibinfo{author}{\bibfnamefont{A.}~\bibnamefont{Kiraz}},
  \bibinfo{author}{\bibfnamefont{M.}~\bibnamefont{Atature}}, \bibnamefont{and}
  \bibinfo{author}{\bibfnamefont{A.}~\bibnamefont{Imamoglu}},
  \bibinfo{journal}{Phys. Rev. A} \textbf{\bibinfo{volume}{69}},
  \bibinfo{pages}{32305} (\bibinfo{year}{2004}).

\bibitem[{\citenamefont{Troiani et~al.}(2006)\citenamefont{Troiani, Perea, and
  Tejedor}}]{troiani06a}
\bibinfo{author}{\bibfnamefont{F.}~\bibnamefont{Troiani}},
  \bibinfo{author}{\bibfnamefont{J.~I.} \bibnamefont{Perea}}, \bibnamefont{and}
  \bibinfo{author}{\bibfnamefont{C.}~\bibnamefont{Tejedor}},
  \bibinfo{journal}{Phys. Rev. B} \textbf{\bibinfo{volume}{73}},
  \bibinfo{pages}{035316} (\bibinfo{year}{2006}).

\bibitem[{\citenamefont{Sheng and Leburton}(2002)}]{sheng02}
\bibinfo{author}{\bibfnamefont{W.}~\bibnamefont{Sheng}} \bibnamefont{and}
  \bibinfo{author}{\bibfnamefont{J.~P.} \bibnamefont{Leburton}},
  \bibinfo{journal}{Appl. Phys. Lett.} \textbf{\bibinfo{volume}{81}},
  \bibinfo{pages}{4449} (\bibinfo{year}{2002}).

\bibitem[{\citenamefont{Bester et~al.}(2004)\citenamefont{Bester, Shumway, and
  Zunger}}]{bester04}
\bibinfo{author}{\bibfnamefont{G.}~\bibnamefont{Bester}},
  \bibinfo{author}{\bibfnamefont{J.}~\bibnamefont{Shumway}}, \bibnamefont{and}
  \bibinfo{author}{\bibfnamefont{A.}~\bibnamefont{Zunger}},
  \bibinfo{journal}{Phys. Rev. Lett.} \textbf{\bibinfo{volume}{93}},
  \bibinfo{pages}{47401} (\bibinfo{year}{2004}).

\bibitem[{not({\natexlab{a}})}]{nota_a2}
\bibinfo{note}{The first excited charged-exciton state 4 has to be included,
  for its energy separation from 3 is typically of a few meV \cite{krenner06},
  comparable to the detunings $\delta_L$ and $\delta_c$.}

\bibitem[{\citenamefont{Hong et~al.}(1987)\citenamefont{Hong, Ou, and
  Mandel}}]{hong}
\bibinfo{author}{\bibfnamefont{C.~K.} \bibnamefont{Hong}},
  \bibinfo{author}{\bibfnamefont{Z.~Y.} \bibnamefont{Ou}}, \bibnamefont{and}
  \bibinfo{author}{\bibfnamefont{L.}~\bibnamefont{Mandel}},
  \bibinfo{journal}{Phys. Rev. Lett.} \textbf{\bibinfo{volume}{59}},
  \bibinfo{pages}{2044} (\bibinfo{year}{1987}).

\bibitem[{not({\natexlab{b}})}]{nota_a1}
\bibinfo{note}{For simplicity, we keep $\omega^A_{34}=\omega^B_{34}= 5\,$meV
  and $\omega^A_{12}=\omega^B_{12}= 25\,$meV. Possible differences between the
  bonding-antibonding splittings of the two AMs can be trivially compensated by
  adjusting $\omega_L^A- \omega_L^B$.}

\bibitem[{\citenamefont{Bertoni et~al.}(2004)\citenamefont{Bertoni, Rontani,
  Goldoni, Troiani, and Molinari}}]{bertoni}
\bibinfo{author}{\bibfnamefont{A.}~\bibnamefont{Bertoni}},
  \bibinfo{author}{\bibfnamefont{M.}~\bibnamefont{Rontani}},
  \bibinfo{author}{\bibfnamefont{G.}~\bibnamefont{Goldoni}},
  \bibinfo{author}{\bibfnamefont{F.}~\bibnamefont{Troiani}}, \bibnamefont{and}
  \bibinfo{author}{\bibfnamefont{E.}~\bibnamefont{Molinari}},
  \bibinfo{journal}{Appl. Phys. Lett.} \textbf{\bibinfo{volume}{85}},
  \bibinfo{pages}{4729} (\bibinfo{year}{2004}).

\bibitem[{\citenamefont{Troiani et~al.}(2007)\citenamefont{Troiani, Wilson-Rae,
  and Tejedor}}]{troiani07}
\bibinfo{author}{\bibfnamefont{F.}~\bibnamefont{Troiani}},
  \bibinfo{author}{\bibfnamefont{I.}~\bibnamefont{Wilson-Rae}},
  \bibnamefont{and} \bibinfo{author}{\bibfnamefont{C.}~\bibnamefont{Tejedor}},
  \bibinfo{journal}{Appl. Phys. Lett.} \textbf{\bibinfo{volume}{90}},
  \bibinfo{pages}{144103} (\bibinfo{year}{2007}).

\bibitem[{\citenamefont{Krenner et~al.}(2006)\citenamefont{Krenner, Clark,
  Nakaoka, Bichler, Scheurer, Abstreiter, and Finley}}]{krenner06}
\bibinfo{author}{\bibfnamefont{H.~J.} \bibnamefont{Krenner}},
  \bibinfo{author}{\bibfnamefont{E.~C.} \bibnamefont{Clark}},
  \bibinfo{author}{\bibfnamefont{T.}~\bibnamefont{Nakaoka}},
  \bibinfo{author}{\bibfnamefont{M.}~\bibnamefont{Bichler}},
  \bibinfo{author}{\bibfnamefont{C.}~\bibnamefont{Scheurer}},
  \bibinfo{author}{\bibfnamefont{G.}~\bibnamefont{Abstreiter}},
  \bibnamefont{and} \bibinfo{author}{\bibfnamefont{J.~J.}
  \bibnamefont{Finley}}, \bibinfo{journal}{Phys. Rev. Lett.}
  \textbf{\bibinfo{volume}{97}}, \bibinfo{pages}{076403}
  (\bibinfo{year}{2006}).

\bibitem[{\citenamefont{Fern\'ee et~al.}(2007)\citenamefont{Fern\'ee,
  Rubinsztein-Dunlop, and Milburn}}]{fernee07}
\bibinfo{author}{\bibfnamefont{M.~J.} \bibnamefont{Fern\'ee}},
  \bibinfo{author}{\bibfnamefont{H.}~\bibnamefont{Rubinsztein-Dunlop}},
  \bibnamefont{and} \bibinfo{author}{\bibfnamefont{G.~J.}
  \bibnamefont{Milburn}}, \bibinfo{journal}{Phys. Rev. A}
  \textbf{\bibinfo{volume}{75}}, \bibinfo{pages}{043815}
  (\bibinfo{year}{2007}).

\bibitem[{\citenamefont{Flindt et~al.}(2007)\citenamefont{Flindt, Sorensen,
  Lukin, and Taylor}}]{flindt}
\bibinfo{author}{\bibfnamefont{C.}~\bibnamefont{Flindt}},
  \bibinfo{author}{\bibfnamefont{A.~S.} \bibnamefont{Sorensen}},
  \bibinfo{author}{\bibfnamefont{M.~D.} \bibnamefont{Lukin}}, \bibnamefont{and}
  \bibinfo{author}{\bibfnamefont{J.~M.} \bibnamefont{Taylor}},
  \bibinfo{journal}{Phys. Rev. Lett.} \textbf{\bibinfo{volume}{98}},
  \bibinfo{pages}{240501} (\bibinfo{year}{2007}).

\end{thebibliography}
\end{document}